\documentclass[showpacs,amssymb,twocolumn,aps,nofootinbib]{revtex4}
\usepackage{amsmath}
\usepackage{amstext}
\usepackage{amsopn}
\usepackage{amsfonts}
\usepackage{amssymb}
\usepackage{bbm}
\usepackage{accents}
\usepackage{empheq}
\usepackage{graphicx}
\usepackage{epsf}
\usepackage{graphics}
\usepackage[latin1]{inputenc}
\def\sc{\scriptstyle}

\def\sl{\!\!\!\slash}

\begin{document}

\title{On the gauge independence of the fermion pole mass}

\author{Ashok K. Das,$^{a,b}$ R. R. Francisco$^{c}$ and J. Frenkel$^{c}$ \footnote{$\ $ e-mail: das@pas.rochester.edu,  jfrenkel@fma.if.usp.br}}
\affiliation{$^a$ Department of Physics and Astronomy, University of Rochester, Rochester, NY 14627-0171, USA}
\affiliation{$^b$ Saha Institute of Nuclear Physics, 1/AF Bidhannagar, Calcutta 700064, INDIA}
\affiliation{$^{c}$ Instituto de Física, Universidade de São Paulo, 05508-090, São Paulo, SP, BRAZIL}

\begin{abstract}
We study the question of complete gauge independence of the fermion pole mass by choosing a general class of gauge fixing which interpolates between the covariant, the axial and the Coulomb gauges for different values of the gauge fixing parameters. We derive the Nielsen identity describing the gauge parameter variation of the fermion two point function in this general class of gauges. Furthermore, we relate the denominator of the fermion propagator to the two point function. This then allows us to study directly the gauge parameter dependence of the denominator of the propagator using the Nielsen identity for the two point function. This leads to a simple proof that, when infrared divergences and mass shell singularities are not present at the pole, the fermion pole mass is gauge independent, in the complete sense, to all orders in perturbation theory. Namely, the pole is not only independent of the gauge fixing parameters, but has also the same value in both covariant and non-covariant gauges.
\end{abstract}


\maketitle

\section{Introduction}

In a relativistic field theory the mass of a particle is given by the Casimir (operator) relation 
\begin{equation}
P_{\mu}P^{\mu} - M_{p}^{2} \mathbbm{1} = 0,\label{intro_1}
\end{equation} 
acting on the one particle state. In an interacting theory, this is computationally quite involved. Instead, therefore, one defines the physical mass of the particle to be given by the pole of its propagator. This follows because the dynamical equation for a relativistic field (free equation after taking quantum corrections into account) is expected to encode the Casimir relation \eqref{intro_1} in some form and, therefore, the zero of the two point function or the pole of the propagator defines the physical mass of the particle. This is computationally much simpler and the mass of a particle can be calculated in a straightforward manner order by order in perturbation theory.

In a theory of fermions interacting with a gauge field, to do any calculaion in perturbation theory, one needs to choose a gauge. Normally one chooses a class of covariant gauge fixing by adding a term to the Lagrangian density of the form (say, in an Abelian theory)
\begin{equation}
{\cal L}_{\rm\sc GF} = - \frac{1}{2\xi}\, (\partial\cdot A)^{2},\label{intro_2}
\end{equation}
which maintains manifest covariance in the intermediate steps of any calculation. Here $\xi$ denotes a gauge fixing parameter. For $\xi=1$ the gauge is known as the Feynman gauge while, for $\xi=0$, it is called the Landau gauge and so on. However, there are also other possible choices of gauge fixing, generically known as non-covariant gauges. Here the choice of gauge fixing depends on a direction conventionally denoted by $n^{\mu}$ (we will discuss this in more detail in sections {\bf 2} and {\bf 3}). The class of generalized axial gauges and Coulomb gauges belong to this class and are also known as physical gauges. In these gauges (for example, in the axial gauges) properties such as asymptotic freedom (in a non-Abelian gauge theory) can be seen in a very simple manner since ghost particles decouple.

As a result of gauge fixing in a gauge theory, the two point function and the propagator (which is a Green's function) of a fermion interacting with a gauge field become gauge dependent. The gauge dependence has two distinct sources, namely, the propagator depends on the gauge fixing parameter (for example, $\xi$ in covariant gauges) and second, its actual form depends on the class of gauge fixing chosen (for example, whether it is the class of  covariant or axial or Coulomb gauges). Nevertheless, one expects that the pole of the propagator should be independent of gauge since it defines a physical quantity (mass). This gauge independence should be complete in the sense that the  pole mass should not only be gauge parameter independent within a class of chosen gauges (say, in covariant gauges), but should also have the same value in different classes of gauges (covariant or non-covariant).

It is indeed important to demonstrate that this expectation is true and this question has been studied within the class of covariant gauges from several points of view \cite{tarrach,johnston,reinders,gray,brown,brecken,kronfeld}. For example, in \cite{tarrach} it was shown in covariant gauges, both in QED and QCD, that the fermion mass is gauge parameter ($\xi$) independent up to two loops in perturbation theory. In \cite{kronfeld} this argument was extended in covariant gauges to show that, in these theories, the fermion pole mass is infrared finite and gauge parameter independent to all orders in perturbation theory. The most direct way to study the gauge dependence of any Green's function or amplitude is through the Nielsen identities \cite{nielsen, das} which follow from the BRST invariance \cite{brst} of the theory. References \cite{johnston,brecken}, in particular, used this approach (again in covariant gauges) to address the question of the gauge parameter independence of the fermion pole mass to all orders in perturbation theory by studying the gauge parameter variation of the fermion two point function.

These studies are all very important even though all of them are carried out within the given class of covariant gauges in \eqref{intro_1}. The gauge parameter independence shows, for example, that the fermion pole mass is the same in the Feynman gauge ($\xi=1$) as in the Landau gauge ($\xi=0$). However, these studies do not address the second source of gauge dependence, namely, one cannot conclude from these studies that the pole mass in the covariant gauges is the same as in non-covariant gauges like the axial gauges or the Coulomb gauges. This still remained to be demonstrated. 

On the other side, there was very little systematic study of the question of gauge independence of the fermion pole mass in non-covariant gauges \cite{kummer, frenkel, leibbrandt, fradkin,bernstein,andrasi}. This is a consequence of the fact that when there is an additional structure ($n^{\mu}$) present, the fermion two point function has a more complicated structure (than in covariant gauges) and extracting the mass from a study of the fermion two point function is nontrivial. In fact, since in this case amplitudes depend both on the momentum as well as the longitudinal component of the momentum (with respect to $n^{\mu}$) of the particle independently, it was an open question as to whether a consistent mass can even be defined in such theories (we will discuss these issues in more detail in the next section).

In a recent brief communication \cite{plb}, using Nielsen identities we gave a simple proof of the complete gauge independence of the fermion pole mass in theories without any infrared divergence and mass shell singularities. This proof shows that the pole mass is gauge parameter independent in both covariant as well as non-covariant gauges and also shows that it is the same in different classes of gauges. This is achieved with three basic ingredients. First, we choose a general class of gauges which interpolate between the covariant and non-covariant gauges (axial and Coulomb) depending on the choice of the gauge fixing parameters \cite{taylor}. We derive the Nielsen identity for the gauge parameter variation of the fermion two point function in such a theory. Finally, since extracting the mass from the two point function is problematic in such a gauge, we study directly the gauge parameter variation of the denominator of the propagator by relating it to the fermion two point function (and using the Nielsen identity).

In this long paper, we give details of our analysis and discuss various other aspects associated with this question. In section {\bf 2}, we describe the generalized axial gauge fixing (in QED) in detail as well as the structure of the fermion self-energy (in the presence of an additional direction $n^{\mu}$) commonly assumed  for such a theory. Using charge conjugation invariance in gauge theories, we show that this structure  simplifies. The conventional definition of the fermion mass, in such theories, is taken directly from the properties of the fermion two point function in covariant gauges. We show that such a definition becomes gauge parameter dependent and does not coincide with the pole of the propagator starting at two loops. This is shown in two ways, first in a completely algebraic manner starting from the definition of the conventional mass taken for such theories and comparing it with the zero of the denominator of the propagator and second from an explicit two loop calculation. This makes clear that the standard definition of the mass taken from studies in covariant gauges does not work in non-covariant gauges and one should really look at the zero of the denominator (of the propagator) to determine the mass. In this section, we also relate the denominator of the fermion propagator to the two point function for later use in showing the gauge independence of the pole of the propagator. In section {\bf 3}, we discuss in detail the choice of an interpolating gauge (still in QED) and derive the Nielsen identity describing the gauge parameter variation of the fermion two point function. In section {\bf 4} we study directly the gauge parameter variation of the denominator of the propagator using the Nielsen identity for the fermion two point function as well as the relation between the denominator and the two point function. The proof of complete gauge independence then follows in a simple manner. In this section, we also show  explicitly the gauge parameter independence of the zero of the denominator (pole of the propagator) up to two loops in the generalized axial gauges. The paper concludes with a short summary of our results and the derivation of the Nielsen identities in QCD in the interpolating gauge is described in the appendix.

\section{Axial gauges}

As we have already mentioned, non-covariant gauges are gauges where the gauge fixing uses an additional direction, $n^{\mu}$, to fix the gauge. (We recall that covariant gauges use only the covariant gradient vector to fix the gauge.)  For example, one may fix the gauge using the longitudinal component of the gradient along the given direction. Such gauges are known as generalized axial gauges. One may alternatively use the transverse component of the gradient (to $n^{\mu}$) to fix the gauge. In particular, if the direction $n^{\mu}$ is timelike, then such gauges are known as generalized Coulomb gauges. Non-covariant gauges are also known as physical gauges.  In generalized axial gauges, the gauge fixing Lagrangian density for QED is given by
\begin{equation}
{\cal L}_{\sc\rm GF} = - \frac{\beta^{2}}{2} \left(\partial_{\sc L}\cdot A\right)^{2},\label{axial_0}
\end{equation}
where $\beta$ is the (constant) gauge fixing parameter and the longitudinal component of the gradient along $n^{\mu}$ is given by
\begin{equation}
\partial^{\mu}_{\sc L} = \frac{(n\cdot \partial) n^{\mu}}{n^{2}}.\label{axial_0a}
\end{equation}
For simplicity of discussions we assume that $n^{2}\neq 0$. With such a gauge fixing, the photon propagator has the form \cite{frenkel}
\begin{widetext}
\begin{equation}
D_{\mu\nu} (p) = - \frac{1}{p^{2}}\left[\eta_{\mu\nu} - \frac{n_{\mu}p_{\nu} + n_{\nu}p_{\mu}}{n\cdot p} + \frac{p_{\mu}p_{\nu}n^{2}}{(n\cdot p)^{2}}\left(1 + \frac{n^{2} p^{2}}{\beta^{2} (n\cdot p)^{2}}\right)\right].\label{axial_0b}
\end{equation}
\end{widetext}
It is worth noting here that the terms inside the square brackets are dimensionless and of order zero in both $n$ and $p$. Furthermore, the propagator, in addition to being symmetric in $p^{\mu}$, is also invariant under $n^{\mu}\leftrightarrow - n^{\mu}$. This symmetry is quite important as we will see in a short while. (The homogeneous axial gauge is obtained in the limit $\beta\rightarrow \infty$.) In axial gauges, the pole at $p^{2} = 0$ is treated, as usual, with the Feynman prescription while the poles at $n\cdot p =0$ are handled with the principal value (PV) prescription. The other thing to emphasize is that the photon propagator has a part (the first term) that coincides with the propagator in the Feynman gauge and is independent of the gauge fixing parameters $\beta, n^{\mu}$ while the rest of the terms depend manifestly on these parameters. So, in calculations studying gauge independence, sometimes it is useful to write the photon propagator \eqref{axial_0b} in the generalized axial gauge as
\begin{equation}
D_{\mu\nu} = D_{\mu\nu}^{(Feynman)} + \widetilde{D}_{\mu\nu} (n, \beta).\label{axial_0c}
\end{equation}

In this section, we will start by recapitulating what is already known about the fermion mass in axial gauges. Let us recall that in covariant gauges the general structure of the fermion self-energy (in a massive theory) is simpler and has the form
\begin{equation}
\Sigma^{(c)} (p) = m A + B p\sl,\label{axial_1}
\end{equation}
where the coefficients $A, B$ are dimensionless functions of the Lorentz invariants $(p^{2},m)$ (in addition to the gauge fixing parameter in a gauge theory). In non-covariant gauges, however, because of the presence of  a nontrivial direction vector $n^{\mu}$, the structure of the fermion self-energy becomes a bit more involved and is conventionally parameterized in the form \cite{pagels, konetschny}
\begin{equation}
\Sigma^{\rm (nc)} (p) = m A + B p\sl + C p\sl_{\sc L} + \frac{m D}{p^{2}_{\sc L}}\left(p\sl_{\sc L} p\sl - p\sl p\sl_{\sc L}\right),\label{axial_2}
\end{equation}
where $p^{\mu}_{\sc L}$ denotes the component of the four momentum $p^{\mu}$ along the given direction $n^{\mu}$. We note here that given a direction vector $n^{\mu}$, any other vector can be decomposed into components parallel and perpendicular to this vector. For example, the momentum vector of the fermion can now have components
\begin{equation}
p^{\mu}_{\sc L} = \frac{(n\cdot p)}{n^{2}}\,n^{\mu},\quad p^{\mu}_{\sc T} = p^{\mu} - p^{\mu}_{\sc L}.\label{axial_3}
\end{equation}
Therefore, we can now construct three Lorentz invariants $p^{2}, p^{2}_{\sc L}$ and $p^{2}_{\sc T}$. However, not all three of these will be independent, rather they will be related as $p^{2} = p^{2}_{\sc L} + p^{2}_{\sc T}$. As a result, only two of them are independent and conventionally, one chooses $p^{\mu}, p^{\mu}_{\sc L}$ as independent vectors which is reflected in \eqref{axial_2}. The coefficients $A, B, C$ and $D$ in \eqref{axial_2} are, in general, dimensionless functions of $(p^{2},p^{2}_{\sc L}, n^{2}, m)$. 

One can understand the rationale behind the structures in \eqref{axial_1} and \eqref{axial_2} simply as follows. A complete basis for matrices in the Dirac space is given by the matrices (generators of the Clifford algebra) $\mathbbm{1}, \gamma^{\mu}, \gamma^{[\mu}\gamma^{\nu]}, \gamma^{[\mu}\gamma^{\nu}\gamma^{\lambda]},\cdots$. Here the square bracket implies anti-symmetrization in the indices. When there is only one independent vector $p^{\mu}$ (as in covariant gauges), the most general structure for a Dirac matrix (without any Lorentz index), in a parity conserving theory, is given by \eqref{axial_1}. On the other hand, if there are two nontrivial independent vectors $p^{\mu}, p^{\mu}_{\sc L}$, it allows for more terms in the expansion in the basis leading to \eqref{axial_2} (in a parity conserving theory). It is clear that not having an additional direction $n^{\mu}$ (or $p^{\mu}_{\sc L}$) is equivalent to having the coefficients $C=D=0$ in \eqref{axial_2} and in this case, the structure reduces to \eqref{axial_1}. 

The general structure of the self-energy in \eqref{axial_2} follows only from the simple structures of the basis of the Clifford algebra and this form has been conventionally used in all earlier discussions. However, in a gauge theory like QED, there are other symmetries which may limit the number of terms in \eqref{axial_2}. One such symmetry is the discrete charge conjugation symmetry which requires that the fermion two point function satisfies
\begin{equation}
{\cal C} \left(S^{-1} (p)\right)^{T} {\cal C}^{-1} = S^{-1} (-p),\label{axial_4}
\end{equation}
where $T$ denotes the transpose of a matrix and  ${\cal C}$ denotes the charge conjugation matrix defined by (we distinguish this from the coefficient $C$ in \eqref{axial_2} with a curly letter)
\begin{equation}
{\cal C}^{-1} \gamma^{\mu} {\cal C} = - \left(\gamma^{\mu}\right)^{T},\quad {\cal C} \left(\gamma^{\mu}\right)^{T} {\cal C}^{-1} = - \gamma^{\mu}.\label{axial_5}
\end{equation}
Requiring charge conjugation invariance, namely, \eqref{axial_4} for
\begin{equation}
S^{-1}_{(nc)} (p) = p\sl - m - \Sigma^{(nc)} (p),\label{axial_6}
\end{equation}
leads to
\begin{align}
A (-p) & = A(p),&\quad &B (-p) = B (p),\notag\\
C(-p) & = C(p),&\quad &D(-p) = - D(p).\label{axial_7}
\end{align}
The last condition on the coefficient $D$ can be satisfied only if it depends on $p^{\mu}$ through an odd power of the Lorentz invariant $n\cdot p$. However, this would violate the symmetry under $n^{\mu}\leftrightarrow -n^{\mu}$ present in the photon propagator in \eqref{axial_0b} (the $n^{\mu}$ dependence in the self-energy arises through the photon propagator). Therefore, we see that charge conjugation invariance in gauge theories like QED requires that
\begin{equation}
D = 0,\label{axial_8}
\end{equation}
We have verified explicitly that \eqref{axial_8} is true at one loop in $n$-dimensions. However, the charge conjugation symmetry restricts this coefficient to vanish identically to all orders. As a result, the general structure of the fermion self-energy is, in fact, simpler than conventionally assumed and has the form
\begin{equation}
\Sigma^{\rm (nc)} (p) = m A + B p\sl + C p\sl_{\sc L},\label{axial_9}
\end{equation}
with the coefficients $A, B, C$ depending only on $(p^{2}, p_{\sc L}^{2}, m, n^{2})$ (as well as the gauge fixing parameter $\beta$). The coefficients $A, B, C$ get contributions from loop diagrams at every order and \eqref{axial_9} represents the self-energy to all orders.

\subsection{Determining the mass}

Let us recapitulate what is done in covariant gauges where things are simpler. The fermion two point function satisfies the equation
\begin{align}
S^{-1}_{(c)}\, u\Big|_{p^{2}=M_{(c)}^{2}} & = (p\sl -m - \Sigma^{(c)}) u\Big|_{p^{2}=M_{(c)}^{2}}\notag\\
& = (p\sl - M_{(c)}) u\Big|_{p^{2}=M_{(c)}^{2}} = 0,\label{axial_10}
\end{align}
where $M_{(c)}$ corresponds to the pole of the propagator (or the vanishing of the determinant). It follows now that
\begin{equation}
\overline{u}\, S^{-1}_{(c)}\, u\Big|_{p^{2}=M_{(c)}^{2}} = \overline{u}\left(p\sl -M_{(c)}\right)u\Big|_{p^{2}= M_{(c)}^{2}} = 0,\label{axial_11}
\end{equation}
which allows us to identify
\begin{equation}
M_{(c)} = m + \overline{u}\,\Sigma^{(c)}\, u\Big|_{p^{2} = M_{(c)}^{2}}.\label{axial_12}
\end{equation}

The dependence on an additional direction $n^{\mu}$, on the other hand, makes the extraction of the physical mass more complicated. Conventionally, one assumes that the mass, $p^{2} = \widetilde{M}^{2}$, can also be obtained, as in covariant gauges (see \eqref{axial_11}), from the relation
\begin{align}
\overline{u} S^{-1}_{\rm (nc)} (p) u\Big|_{p^{2}=\widetilde{M}^{2}} & = \overline{u}(p\sl - m - \Sigma^{(nc)}) u\Big|_{p^{2}=\widetilde{M}^{2}}\notag\\
& = \overline{u} (p\sl -\widetilde{M})u\Big|_{p^{2}=\widetilde{M}^{2}}= 0.\label{axial_13}
\end{align}
This would then determine, as in \eqref{axial_12}, that the mass $\widetilde{M}$ is given by 
\begin{equation}
\widetilde{M} = m + \overline{u}\,\Sigma^{\rm (nc)}\, u\Big|_{p^{2}=\widetilde{M}^{2}}.\label{axial_14}
\end{equation}
In fact, this mass can be evaluated from the fermion self-energy order by order and at one loop it turns out  \cite{konetschny} that $\widetilde{M}$ is gauge parameter independent (independent of $\beta, n^{\mu}$).  This one loop  calculation has led to the expectation that the mass $\widetilde{M}$ determined from \eqref{axial_14} is gauge parameter independent to all orders as in covariant gauges and should correspond to the pole of the propagator. However, this expectation fails beginning at two loops and we will show this, in the following, in two different ways. First, we will show algebraically that $\widetilde{M}$ does not correspond to the pole of the fermion propagator and the difference between the two, starting at two loops, is manifestly gauge parameter dependent ($n^{\mu}$ dependent). Second, we will demonstrate through an explicit calculation of the fermion self-energy at two loops that $\overline{u}\,\Sigma^{\rm (nc)}\,u$ is manifestly gauge dependent at $p^{2} = \widetilde{M}^{2}$ (consistent with the algebraic result) so that $\widetilde{M} = \widetilde{M} (n)$ if \eqref{axial_14} holds. (The explicit calculation shows that $\widetilde{M}$ does not depend on the parameter $\beta$ at two loops.) The reason for this unexpected behavior can be traced to the fact that in non-covariant gauges where there is an additional structure ($n^{\mu}$) present, $\overline{u}\, S^{-1} (p)u\big|=0$ does not imply that $S^{-1} (p)u\big|=0$ where the restriction implies evaluating this at $p^{2}=\widetilde{M}^{2}$ (of course, the converse is always true). As a result $\widetilde{M}$ determined by relation \eqref{axial_14} does not represent the pole of the propagator beyond the one loop order.  To demonstrate all this, we need the form of the fermion propagator which we derive in the next subsection.

\subsection{Fermion propagator}

Let us recall from the form of the fermion self-energy in \eqref{axial_9} that in the (non-covariant) axial gauge the fermion two point function can be written as
\begin{equation}
S^{-1}(p) = {\cal A} - {\cal B} \mathbbm{1},\label{axial_15}
\end{equation}
where ${\cal A}$ is a nontrivial matrix while ${\cal B}$ is a scalar of the forms
\begin{equation}
{\cal A} = (1-B) p\sl - C p\sl_{\sc L},\quad {\cal B} = m (1+A).\label{axial_16}
\end{equation}
It is straightforward to check from \eqref{axial_16} (and the properties of the Dirac gamma matrices as well as the definition of the longitudinal momentum) that 
\begin{equation}
{\cal A}^{2} = \left( (1-B)^{2} p^{2} + (C^{2} - 2C (1-B)) p_{\sc L}^{2}\right) \mathbbm{1}.\label{axial_17}
\end{equation}
In such a case, the propagator (up to a factor $i$) can be obtained in a simple manner to have the form
\begin{equation}
S (p) = \frac{\cal N}{\cal D},\label{axial_18}
\end{equation}
where (the identity matrix in the denominator is not relevant and we have used \eqref{axial_4})
\begin{widetext}
\begin{align}
{\cal N} & = {\cal A} + {\cal B} \mathbbm{1} = - {\cal C} (S^{-1} (p))^{T} {\cal C}^{-1} = - S^{-1} (-p),\label{axial_19}\\ 
{\cal D} & = {\cal A}^{2} - {\cal B}^{2} = (1-B)^{2} p^{2} + (C^{2} - 2C (1-B))p_{\sc L}^{2} - m^{2} (1+A)^{2}.\label{axial_20}
\end{align}
\end{widetext}

The pole of the propagator is determined from the zero of the denominator ${\cal D}$ and since the denominator depends on both $p^{2}$ and $p_{\sc L}^{2}$, it is not clear {\em a priori} whether a mass can even be defined. (This is quite different from covariant gauges.) For example, suppose the zero of the denominator occurs at $p^{2} = M_{p}^{2}$, then from \eqref{axial_20} we obtain
\begin{equation}
M_{p}^{2} = \frac{[m^{2}(1+A)^{2} - (C^{2} - 2C(1-B))p_{\sc L}^{2}]}{(1-B)^{2}}\Big|_{p^{2}=M_{p}^{2}},\label{axial_21}
\end{equation}
so that, in general, the location of the pole of the propagator seems to depend on the longitudinal momentum $p_{\sc L}$ and, therefore, cannot define a mass (unless all the dependence on the longitudinal momentum  $p_{\sc L}$ cancels out). As we will show in section {\bf 4} using the Nielsen identity, this is indeed the case and $M_{p}$ is, in fact, independent of the gauge parameters $\beta$ as well as $n^{\mu}$. For later use, we note from \eqref{axial_18}-\eqref{axial_20} that we can write
\begin{align}
{\cal D} \mathbbm{1} & = - S^{-1}(p) {\cal C} (S^{-1}(p))^{T} {\cal C}^{-1}\notag\\
& = - S^{-1} (p) S^{-1} (-p).\label{axial_22}
\end{align}
Alternatively, taking the matrix trace in \eqref{axial_22} we determine the scalar denominator of the propagator to be given by
\begin{align}
{\cal D} & = - \frac{1}{2^{[n/2]}}\, \text{Tr}  \left(S^{-1}(p){\cal C} (S^{-1} (p))^{T} {\cal C}^{-1}\right)\notag\\
& = - \frac{1}{2^{[n/2]}}\,\text{Tr}\left(S^{-1}(p) S^{-1} (-p)\right).\label{axial_23}
\end{align}
Here $n$ denotes the dimensionality of space-time, $[n/2]$ represents the integer part of $n/2$ and we have used the fact that, in $n$-dimensions, the Dirac matrices are $2^{[n/2]}\times 2^{[n/2]}$ matrices. (For completeness, note from \eqref{axial_15} and \eqref{axial_19} that $S^{-1} (p)$ and $S^{-1} (-p)$ commute.)
We will show in section {\bf 4} that the pole of the propagator given in \eqref{axial_21} is gauge independent. For the moment, we only note that since $A,B,C$ receive contributions at various loops, the pole \lq\lq mass" $M_{p}$ can also be expanded in powers of loops as
\begin{equation}
M_{p} = M_{p}^{(0)} + M_{p}^{(1)} + M_{p}^{(2)} + \cdots,\label{axial_24} 
\end{equation}
with the tree level mass identified with $M_{p}^{(0)} = m$ (namely, at tree level $A^{(0)}=B^{(0)}=C^{(0)}=0$).

On the other hand, $\widetilde{M}$ is defined in \eqref{axial_14} and to evaluate this we need an expression for $\overline{u} \gamma^{\mu}u$. In covariant theories with momentum as the only available four vector, this is normally determined to be $\overline{u} \gamma^{\mu} u\big|_{p^{2}=M^{2}} = \frac{p^{\mu}}{M}$ (so that we have $\overline{u}\,p\sl\, u\big|_{p^{2}=M^{2}} = M$). When there is an additional structure ($n^{\mu}$ or $p_{\sc L}^{\mu}$) present, this generalizes to
\begin{equation}
\overline{u} \gamma^{\mu} u\Big|_{p^{2}=\widetilde{M}^{2}} = (1-a)\,\frac{p^{\mu}}{\widetilde{M}} + a \widetilde{M}\,\frac{p_{\sc L}^{\mu}}{p_{\sc L}^{2}}.\label{axial_25}
\end{equation}
Here $a$ is a nontrivial parameter, in general, beginning at one loop and beyond since the additional structures arise in the self-energy only in that order. It follows from \eqref{axial_25} that (see also \eqref{axial_13})
\begin{align}
& \overline{u}\, p\sl\, u\Big|_{p^{2}=\widetilde{M}^{2}} = \widetilde{M},\notag\\
& \overline{u}\, p\sl_{\sc L}\, u\Big|_{p^{2}=\widetilde{M}^{2}} = \frac{p_{\sc L}^{2}}{\widetilde{M}} - a \widetilde{M}\left(\frac{p_{\sc L}^{2}}{\widetilde{M}^{2}} -1\right).\label{axial_26}
\end{align}
Using \eqref{axial_26}, we can now evaluate $\widetilde{M}$ from the definition in \eqref{axial_14} leading to
\begin{widetext}
\begin{equation}
\widetilde{M}^{2} = \frac{1}{(1-B)^{2}}\left[m^{2}(1+A)^{2} - C^{2} \left(\frac{p_{\sc L}^{2}}{\widetilde{M}} - a\widetilde{M}\left(\frac{p_{\sc L}^{2}}{\widetilde{M}^{2}} -1\right)\right)^{2} + 2C(1-B) \left(p_{\sc L}^{2} - a \widetilde{M}^{2} \left(\frac{p_{\sc L}^{2}}{\widetilde{M}^{2}} -1\right)\right)\right]_{p^{2}=\widetilde{M}^{2}}.\label{axial_27}
\end{equation}
\end{widetext}
Comparing \eqref{axial_21} and \eqref{axial_27}, it is clear that $M_{p}$ and $\widetilde{M}$ are, in general, different, but it is not clear at what order the difference arises. To that end, we note that we can also expand
\begin{equation}
\widetilde{M} = \widetilde{M}^{(0)} + \widetilde{M}^{(1)} + \widetilde{M}^{(2)} + \cdots.\label{axial_28}
\end{equation}
From \eqref{axial_21} and \eqref{axial_27} we can now determine
\begin{align}
\widetilde{M}^{(0)} & = M_{p}^{(0)} = m,\notag\\
\widetilde{M}^{(1)} & = M_{p}^{(1)} = m (A^{(1)} + B^{(1)}) + C^{(1)}\frac{p_{\sc L}^{2}}{m}\Big|_{p^{2}=m^{2}},\notag\\
\widetilde{M}^{(2)} & = M_{p}^{(2)} - \left(\!(C^{(1)})^{2}\frac{p_{\sc L}^{2}}{2m} + ma^{(1)}C^{(1)}\!\right)\!\!\left(\frac{p_{\sc L}^{2}}{m^{2}}-1\right)_{p^{2}=m^{2}},\label{axial_29}
\end{align}
and so on. Namely, $\widetilde{M}$ defined by \eqref{axial_14} coincides with the pole of the propagator only up to one loop, but this equivalence fails beginning at two loops. If $M_{p}$ is gauge independent, as we will see in section {\bf 4}, then \eqref{axial_29} shows that $\widetilde{M}$ becomes manifestly gauge dependent beginning at two loops. However, note from \eqref{axial_29} that in the special gauge $n^{\mu}\parallel p^{\mu}$ or $p_{\sc L}^{\mu} = p^{\mu}$, the difference vanishes. This is, in fact, not just an accident at two loops, rather it holds to all orders which can be seen as follows. In the gauge $p_{\sc L}^{\mu} = p_{\mu}$, we note that \eqref{axial_21} and \eqref{axial_27} reduce to the same equation and, therefore, coincide to all orders. Another way of seeing this is to note that in the special gauge $p_{\sc L}^{\mu} = p^{\mu}$, there is no additional structure and the fermion two point function has the same structure as in a covariant gauge for which the two equations \eqref{axial_10} and \eqref{axial_11} are equivalent. 

\begin{widetext}
\begin{figure}[ht!]
\begin{center}
\includegraphics[scale=.8]{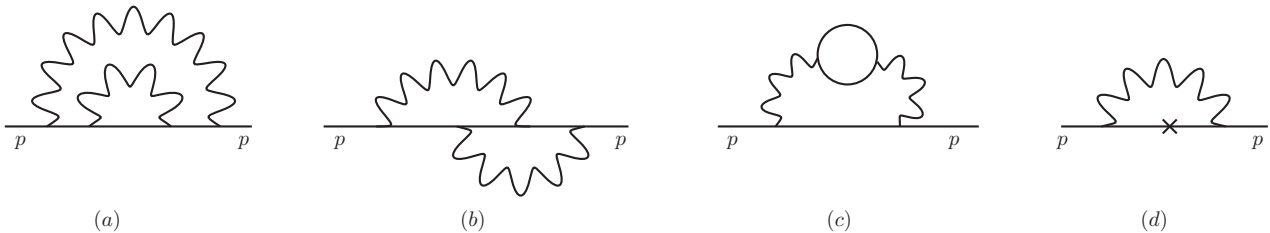}
\end{center}
\caption{Feynman diagrams for the fermion self-energy in QED at two loops.}
\label{fig1}
\end{figure}
\end{widetext}

We can also explicitly evaluate the self-energy through Feynman diagrams and determine $\widetilde{M}$ from there. It is already known from such a calculation that $\widetilde{M}$ is gauge parameter independent (independent of $\beta, n^{\mu}$) at one loop. So, let us concentrate on the contributions at two loops. At this order the relevant Feynman diagrams are given by Fig. \ref{fig1}. 
Here the diagrams $(a), (b), (c)$ denote the standard two loop graphs while the diagram in $(d)$ represents the two loop contribution coming from the one loop mass correction. We note that since the photon self-energy in $(c)$ is transverse, the last term in the photon propagator in \eqref{axial_0b} gives vanishing contribution so that this diagram is independent of $\beta$. In fact, the $\beta$ dependent terms in all the other diagrams identically vanish because of the PV prescription. There are several other cancellations in the $n^{\mu}$ dependent terms involving diagrams $(a), (b), (c), (d)$ and the final result for $\widetilde{M}$ from this explicit calculation takes the factorized  form (in $n$-dimensions)
\begin{widetext}
\begin{equation}
\widetilde{M} = M^{(Feynman)} + \frac{e^{4}}{(2\pi)^{2n}}\,\overline{u} (p) \int \frac{d^{n}q_{1}}{q_{1}^{2} (n\cdot q_{1})}\, n\sl \frac{1}{p\sl + q\sl_{1} - m}\, (p\sl - m) \int \frac{d^{n}q_{2}}{q_{2}^{2} (n\cdot q_{2})}\,\frac{1}{p\sl + q\sl _{2} - m}\, n\sl\, u (p)\Big|_{p^{2}=m^{2}},\label{axial_30}
\end{equation}
\end{widetext}
where $q_{1}, q_{2}$ denote the two internal loop momenta and $M^{(Feynman)}$ is the mass obtained from the first term ($\beta, n^{\mu}$ independent term) in the propagator in \eqref{axial_0b} or \eqref{axial_0c}. Since $M^{(Feynman)}$ is independent of the gauge parameters, it shows explicitly that $\widetilde{M}$ is manifestly gauge dependent at two loops consistent with the algebraic result in \eqref{axial_29}. The reason for this gauge dependence is clear from the structure in \eqref{axial_30}, namely, that it is not the factor $(p\sl - m)$, rather $n\sl$ which occurs at the two ends.  Since $p\sl$ and $n\sl$ do not commute, we cannot move $(p\sl -m)$ move past $n\sl$ to obtain a vanishing result for the second term. However, note that if $n^{\mu}$ is parallel to $p^{\mu}$, then each of the (factorized) integrals can be evaluated to have the form $(a p\sl + b)$ which will commute with $(p\sl -m)$ and, in this case, we can take the factor $(p\sl - m)$ to one end to annihilate the spinor. So, only for this special gauge will $\widetilde{M}$ coincide with $M^{(Feynman)}$ as we have also seen in the algebraic method.

As we have tried to emphasize in this section, the conventionally accepted definition of mass of the fermion $\widetilde{M}$ in non-covariant gauges, \eqref{axial_14}, does not coincide with the pole of the propagator $M_{p}$ beyond one loop and becomes manifestly gauge parameter dependent. However, we are yet to show that the actual pole defines a gauge independent mass. As we have already pointed out in \eqref{axial_21}, it is not clear that we can even define a meaningful mass in non-covariant gauges. In the next two sections, we will show through Nielsen identities that the pole $M_{p}$ indeed defines a mass and is gauge independent in the complete sense.

\section{Nielsen identity in an interpolating gauge}

We know that the fermion self-energy and, therefore, the two point function become gauge dependent once a gauge fixing condition has been chosen and the gauge parameter variation of the fermion two point function is best studied through the Nielsen identities. However, we wish to study the question of gauge independence of the fermion pole mass completely in the sense that we wish to show that the pole mass is not only independent of the gauge fixing parameter within a given class of gauges such as covariant or Coulomb or axial gauges, but also has the same value in all of these gauges. We achieve this in three basic steps, (i) choose a gauge fixing which will interpolate between covariant, Coulomb and axial gauges for different values of the gauge fixing parameters, (ii) derive the Nielsen identity for the gauge parameter variation of the fermion two point function for such a gauge fixing and (iii)  study the gauge parameter variation of the zero of the denominator (of the propagator) \eqref{axial_20},  using the Nielsen identity for the two point function as well as the relation between the denominator and the two point function already discussed in \eqref{axial_23}. In this section, we only take up the first two steps leaving the last to the next section.

\subsection{Choice of an interpolating gauge}

Let us consider massive QED in $n$ space-time dimensions (the generalization to QCD is discussed in the appendix) described by the Lagrangian density
\begin{equation}
{\cal L}_{\rm inv} = - \frac{1}{4}\, F_{\mu\nu}F^{\mu\nu} + \overline{\psi} (i D\!\sl - m) \psi.\label{nielsen_1}
\end{equation}
As the first step, our goal is to study this theory in a general class of gauges which can interpolate between the covariant, the Coulomb and the axial gauges. To that end, we choose a gauge fixing Lagrangian density of the form  \cite{taylor}
\begin{equation}
{\cal L}_{\rm\sc GF} = - \frac{1}{2}\left(\Lambda^{\mu} (\partial) A_{\mu}\right)^{2},\label{nielsen_2}
\end{equation}
where
\begin{equation}
\Lambda^{\mu} (\partial) = \alpha \partial^{\mu} + \beta \partial^{\mu}_{\sc L},\quad \partial^{\mu}_{\sc L} = \frac{(n\cdot \partial)}{n^{2}}\,n^{\mu},\label{nielsen_3}
\end{equation}
and $\alpha,\beta$ are arbitrary constant parameters. Unlike usual gauge fixing Lagrangian densities which depend only on one gauge fixing parameter, here the dependence is on three independent parameters $\alpha, \beta$ and $n^{\mu}$ which we denote collectively by
\begin{equation}
\phi_{(a)} = (\alpha, \beta, n^{\mu}).\label{nielsen_4}
\end{equation} 
The reason for this, as we have already emphasized earlier, is that this is an interpolating gauge between various classes of gauges in the sense that 
\begin{align}
\beta & = 0,&  & \text{covariant gauges}\ (\text{with}\ \xi=\alpha^{-2}),\notag\\
\alpha & = 0,& & \text{generalized axial gauges},\notag\\
\beta & = -\alpha, n^{2}>0,& & \text{generalized Coulomb gauges}.\label{nielsen_4a}
\end{align}
So, for different choices of the gauge fixing parameters one can easily go from covariant to axial to Coulomb  gauges (which is our purpose in choosing such a gauge fixing term). Let us also note that, for the purposes of manifest BRST invariance (which is essential for deriving Ward identities and Nielsen identities), the gauge fixing Lagrangian density \eqref{nielsen_2} can also be written with an auxiliary (nondynamical) field $F$ as
\begin{equation}
{\cal L}_{\sc\rm GF} = \frac{1}{2}\, F^{2} + \left(\Lambda^{\mu} (\partial) F\right) A_{\mu}.\label{nielsen_5}
\end{equation}

The ghost Lagrangian density corresponding to this general class of gauge choice, \eqref{nielsen_5}, is given by
\begin{equation}
{\cal L}_{\rm ghost} = \left(\Lambda^{\mu} (\partial) \overline{c}\right) \partial_{\mu}c,\label{nielsen_6}
\end{equation}
and the combined Lagrangian density 
\begin{equation}
{\cal L} = {\cal L}_{\rm inv} + {\cal L}_{\rm\sc GF} + {\cal L}_{\rm ghost},\label{nielsen_6a}
\end{equation} 
is manifestly invariant (with gauge fixing described by \eqref{nielsen_5}) under the standard (nilpotent) BRST transformations of QED \cite{brst,das},
\begin{align}
& \delta A_{\mu} = \omega \partial_{\mu}c, &  & \delta F = 0,\notag\\
& \delta \psi = - ie\omega c \psi, &  & \delta \overline{\psi} = -ie\omega \overline{\psi} c,\notag\\
& \delta c = 0, & & \delta \overline{c} = - \omega F,\label{nielsen_7}
\end{align}
where $\omega$ represents an arbitrary constant Grassmann parameter. For completeness we note here that the photon propagator for this generalized gauge fixing can be easily obtained to have the form \cite{taylor}
\begin{widetext}
\begin{equation}
D_{\mu\nu} (p) = - \frac{1}{p^{2}}\left[\eta_{\mu\nu} - \frac{\Lambda_{\mu} (p) p_{\nu} + \Lambda_{\nu}(p) p_{\mu}}{\Lambda(p)\cdot p} + \frac{p_{\mu}p_{\nu} (\Lambda(p))^{2}}{(\Lambda(p)\cdot p)^{2}}\left(1 + \frac{p^{2}}{(\Lambda(p))^{2}}\right)\right],\label{nielsen_7a}
\end{equation}
\end{widetext}
which can be compared with \eqref{axial_0b}. Furthermore, it is straightforward to check that for different choices of the gauge fixing parameters, as in \eqref{nielsen_4a}, this propagator reduces to the familiar propagators in the three different classes of gauges.

\subsection{Nielsen identities}

To derive the Green's functions of the theory we need to introduce sources for the fundamental fields and to determine the  gauge parameter variations of the effective action and the Green's functions (Nielsen identities), we also need to introduce some other sources and these are described by the Lagrangian density
\begin{align}
{\cal L}_{\rm source} & = J^{\mu}A_{\mu} + JF + i\left(\overline{\chi}\psi - \overline{\psi}\chi\right) + i\left(\overline{\eta} c - \overline{c}\eta\right)\notag\\
& + ie\left(\overline{M} c\psi - \overline{\psi}c M\right) + \left(H_{(\alpha)} (\partial^{\mu}\overline{c}) + H_{(\beta)} (\partial^{\mu}_{\sc L}\overline{c})\right)A_{\mu}\notag\\
& + \beta H_{(n)\,\mu} (N^{\mu\nu}\overline{c})A_{\nu}.\label{nielsen_8}
\end{align}
Here we have identified
\begin{equation}
N^{\mu\nu} = \frac{\partial \partial^{\nu}_{\sc L}}{\partial n_{\mu}} = \frac{(n\cdot \partial)}{n^{2}} \left(\eta^{\mu\nu} + \frac{\partial^{\mu}n^{\nu}}{n\cdot\partial} - \frac{2n^{\mu}n^{\nu}}{n^{2}}\right).\label{nielsen_9}
\end{equation}
Therefore, the total Lagrangian density for the gauge fixed theory (with sources) is given by
\begin{align}
{\cal L}_{\rm\sc TOT} & = {\cal L}_{\rm inv} + {\cal L}_{\rm\sc GF} + {\cal L}_{\rm ghost} + {\cal L}_{\rm source},\notag\\
S_{\rm\sc TOT} & = \int d^{n}x\, {\cal L}_{\rm\sc TOT}.\label{nielsen_10}
\end{align}
As mentioned above, we note from \eqref{nielsen_8} that $(J^{\mu}, J, \overline{\chi}, \chi, \overline{\eta}, \eta)$ correspond respectively to the standard sources for the fields $(A_{\mu}, F, \psi, \overline{\psi}, c, \overline{c})$; the sources $(\overline{M}, M)$ generate respectively the composite BRST variations of $(\psi, \overline{\psi})$ in \eqref{nielsen_7}. In addition, we have introduced three sources
\begin{equation}
H_{(a)} = \left(H_{(\alpha)}, H_{(\beta)}, H_{(n)}^{\mu}\right),\label{nielsen_11}
\end{equation}
such that the BRST variations of these three source terms lead to the gauge variations of ${\cal L}_{\rm inv} + {\cal L}_{\rm\sc GF} + {\cal L}_{\rm ghost}$ with respect to the three parameters $\phi_{(a)}$ defined in \eqref{nielsen_4}. Note here that the additional sources $(\overline{M}, M)$ are commuting sources which carry spinor indices and ghost quantum numbers while the three sources $H_{(a)}$ defined in \eqref{nielsen_11} are anti-commuting and carry ghost quantum numbers (but have no spinor index).

The generating functional for this gauge fixed theory is now given by
\begin{equation}
Z [{\cal J}] = e^{iW[{\cal J}]} = N \int {\cal D}\varphi\, e^{iS_{\rm\sc TOT}},\label{nielsen_12}
\end{equation}
where we have denoted all the field variables and sources generically by $\varphi$ and ${\cal J}$ respectively and $N$ denotes the normalization constant for the path integral. If we make a field redefinition corresponding to the BRST transformations \eqref{nielsen_7}, namely, 
\begin{equation}
\varphi \rightarrow \varphi + \delta_{\sc BRST}\varphi,\label{nielsen_13}
\end{equation}
inside the path integral, the generating functional \eqref{nielsen_12} does not change since the path integral involves integration over all field configurations (alternatively, the generating functional depends only on the sources and not on the fields). As a result, we obtain
\begin{equation}
\delta Z [{\cal J}] = 0 = iN\int {\cal D}\varphi\, (\delta_{\sc BRST} S_{\rm\sc TOT})\, e^{iS_{\rm\sc TOT}}.\label{nielsen_14}
\end{equation}
Only the source terms in $S_{\rm\sc TOT}$ in \eqref{nielsen_10} are not BRST invariant and substituting the variations of these terms in \eqref{nielsen_14} leads to an identity involving the generating functional $W[{\cal J}]$ of the form (we use the convention of left derivatives for anti-commuting variables)
\begin{align}
& N e^{-iW}\int {\cal D}\varphi \int \!\!d^{n}x\left(\!H_{(a)}\frac{\partial {\cal L}_{\rm\sc TOT}}{\partial\phi_{(a)}} + O ((H_{(a)})^{2})\right) e^{iS_{\rm\sc TOT}}\notag\\
& = i\int d^{n}x\left(J^{\mu}\partial_{\mu} \frac{\delta W}{\delta\overline{\eta}} - \overline{\chi} \frac{\delta W}{\delta\overline{M}} + \frac{\delta W}{\delta M} \chi - \frac{\delta W}{\delta J} \eta\right).\label{nielsen_15}
\end{align}
Furthermore, taking the functional derivative with respect to $H_{(a)}(x)$ and setting the sources $H_{(a)}=0$ and then integrating over $\int d^{n}x$, we obtain the Master identity for the generating functional for  connected Green's functions
\begin{widetext}
\begin{equation}
\frac{\partial W}{\partial \phi_{(a)}} = i \int d^{n}x\,d^{n}y\left(J^{\mu}(y)\partial^{(y)}_{\mu} \frac{\delta^{2} W}{\delta H_{(a)}(x)\delta\overline{\eta}(y)} + \overline{\chi}(y) \frac{\delta^{2} W}{\delta\overline{M}(y) \delta H_{(a)}(x)} + \frac{\delta^{2} W}{\delta H_{(a)}(x) \delta M(y)} \chi (y) - \frac{\delta^{2} W}{\delta H_{(a)}(x) \delta J(y)} \eta(y)\right).\label{nielsen_16}
\end{equation}
\end{widetext}
Upon differentiation, this equation leads to the gauge parameter variation of the connected Green's functions of the theory. We emphasize that all the sources $H_{(a)}$ have been set to zero in \eqref{nielsen_16}.

To obtain the gauge parameter variation of the 1PI amplitudes of the theory, we make a Legendre transformation with respect to the standard sources for the dynamical fields of the theory, namely, (here the field variables correspond to the classical fields)
\begin{equation}
\Gamma = W - \int d^{n}x \left(J^{\mu}A_{\mu} + JF + i (\overline{\chi}\psi - \overline{\psi} \chi) + i (\overline{\eta}c - \overline{c}\eta)\right).\label{nielsen_17}
\end{equation}
This leads (from \eqref{nielsen_16}) to the Master identity for the gauge parameter variation of the effective action in the form
\begin{align}
\frac{\partial \Gamma}{\partial\phi_{(a)}} & = \int d^{n}z d^{n}w\left(\frac{\delta\Gamma}{\delta\psi_{\gamma}(w)} \frac{\delta^{2}\Gamma}{\delta \overline{M}_{\gamma} (w) \delta H_{(a)}(z)}\right.\notag\\
&\qquad\left. + \frac{\delta^{2}\Gamma}{\delta H_{(a)}(z) \delta M_{\gamma}(w)} \frac{\delta\Gamma}{\delta \overline{\psi}_{\gamma}(w)}\right),\label{nielsen_18}
\end{align}
where, again, the sources $H_{(a)}$ have been set to zero. By taking derivatives of this identity with respect to various fields and setting all the fields (including $M, \overline{M}$) to zero, we can obtain the gauge parameter variation of any 1PI amplitude in the theory. For example, taking the functional derivative with respect to $\frac{\delta^{2}}{\delta\psi_{\beta}(y)\delta\overline{\psi}_{\alpha}(x)}$ and setting all fields to zero, we obtain the gauge parameter variation of the fermion two point function
\begin{align}
\frac{\partial S^{-1}_{\alpha\beta}(x-y)}{\partial\phi_{(a)}}  & = \int d^{n}z d^{n}w\left({\cal F}^{(a)}_{\alpha\gamma} (x,z,w) S^{-1}_{\gamma\beta} (w-y)\right.\notag\\
&\quad \left. + S^{-1}_{\alpha\gamma}(x-w) {\cal G}^{(a)}_{\gamma\beta} (w,z,y)\right).\label{nielsen_19}
\end{align}
Here we have identified
\begin{align}
S^{-1}_{\alpha\beta} (x-y) & = \frac{\delta^{2}\Gamma}{\delta\psi_{\beta} (y)\delta\overline{\psi}_{\alpha} (x)},\notag\\
{\cal F}^{(a)}_{\alpha\gamma}(x,z,w) & = \frac{\delta^{3}\Gamma}{\delta\overline{\psi}_{\alpha}(x)\delta H_{(a)}(z)\delta M_{\gamma}(w)},\notag\\
{\cal G}^{(a)}_{\gamma\beta} (w,z,y) & = \frac{\delta^{3}\Gamma}{\delta\overline{M}_{\gamma}(w)\delta H_{(a)}(z)\delta\psi_{\beta}(y)},\label{nielsen_20}
\end{align}
with all field variables (including $M, \overline{M}$) set to zero. In momentum space, the Nielsen identity \eqref{nielsen_19} takes the simple form
\begin{equation}
\frac{\partial S^{-1}_{\alpha\beta} (p)}{\partial \phi_{(a)}} = {\cal F}^{(a)}_{\alpha\gamma} (p) S^{-1}_{\gamma\beta} (p) + S^{-1}_{\alpha\gamma} (p) {\cal G}^{(a)}_{\gamma\beta} (p),\label{nielsen_21}
\end{equation}
where we have identified the three point amplitudes  
\begin{align}
{\cal F}^{(a)}_{\alpha\gamma} (p) & = {\cal F}^{(a)}_{\alpha\gamma} (-p, 0,p),\notag\\
{\cal G}^{(a)}_{\gamma\beta} (p) & = {\cal G}^{(a)}_{\gamma\beta} (-p.0.p).\label{nielsen_22}
\end{align}
Equation \eqref{nielsen_21} shows how the fermion two point function changes with respect to the three independent parameters $(\alpha,\beta,n^{\mu})$. We note here that since there are no vertices corresponding to ${\cal F}^{(a)}_{\alpha\beta} (p), {\cal G}^{(a)}_{\alpha\beta} (p)$ at the tree level Lagrangian density in \eqref{nielsen_10}, these amplitudes are nontrivial only at one loop and beyond. Therefore, the Nielsen identity \eqref{nielsen_21} implies that the dependence of the fermion two point function on the gauge fixing parameter arises only at one loop and beyond, as we expect.

\section{Gauge independence of the pole of the fermion propagator}

We are now ready to show the gauge independence of the pole of the fermion propagator. We note that studying the pole of the propagator is equivalent to studying the zero of the denominator ${\cal D}$ of the propagator defined in \eqref{axial_18} and \eqref{axial_20}. Studying directly the gauge parameter independence of the zero of the denominator in the form \eqref{axial_20} (in a generalized interpolating gauge) is more involved than in the covariant gauge because of the presence of an additional structure. There is a much simpler way to do this which we discuss in the following. Let us recall from \eqref{axial_23} that the denominator of the fermion propagator can be related to the fermion two point function as
$$
{\cal D} = - \frac{1}{2^{[n/2]}}\, \text{Tr}  \left( S^{-1} (p) {\cal C} (S^{-1} (p))^{T} {\cal C}^{-1}\right).
$$
Since the Nielsen identity \eqref{nielsen_21} describes how the fermion two point function changes with the change in any of the three gauge parameters, we can now use this to determine the gauge parameter variation of the denominator of the fermion propagator
\begin{align}
& \frac{\partial {\cal D}}{\partial\phi_{(a)}} = - \frac{1}{2^{[n/2]}}\,{\rm Tr}\left(\frac{\partial S^{-1} (p)}{\partial\phi_{(a)}} {\cal C} (S^{-1}(p))^{T} {\cal C}^{-1}\right.\notag\\
& \qquad\qquad\qquad\qquad\left. + S^{-1}(p) {\cal C} \frac{\partial (S^{-1}(p))^{T}}{\partial\phi_{(a)}} {\cal C}^{-1}\right)\notag\\
& = {\cal D}\, {\rm Tr}\left(\!{\cal F}^{(a)}(p) + {\cal G}^{(a)}(p) + {\cal C} \left({\cal F}^{(a)}(p) + {\cal G}^{(a)}(p)\right)^{T}\!{\cal C}^{-1}\!\right)\notag\\
& = 2 {\cal D}\, {\rm Tr} ({\cal F}^{(a)}(p) + {\cal G}^{(a)}(p)).\label{gaugeind_1}
\end{align}
Here we have used \eqref{axial_22}, cyclicity of trace as well as the fact that the trace of the transpose of a matrix coincides with that of the matrix itself. 

Equation \eqref{gaugeind_1} shows explicitly that the denominator depends on the gauge parameters (is gauge dependent) and the gauge parameter variation of the denominator with respect to the three independent parameters is proportional to the denominator itself. 
As a result, it follows that, if the amplitudes $({\cal F}^{(a)} (p), {\cal G}^{(a)}(p))$ are well behaved, the zero of the denominator ${\cal D}$ (which corresponds to the pole of the propagator) is, in fact, gauge parameter independent, namely,
\begin{equation}
\frac{\partial {\cal D}}{\partial \phi_{(a)}}\Big|_{{\cal D}=0} = 0.\label{gaugeind_2}
\end{equation}
This would, in fact, be the case when the theory does not have any infrared divergences or mass shell singularities at the pole of the propagator. Actually, this turns out to be the case in perturbative QED (and in QCD) in $4$ space-time dimensions \cite{kronfeld}. On the other hand, in lower dimensions, $n < 4$, such singularities can be present in the amplitudes $({\cal F}^{(a)} (p), {\cal G}^{(a)} (p))$ so that the pole of the propagator may become gauge dependent. This has been explicitly studied in the massive Schwinger model (massive QED in $1+1$ dimensions) in \cite{schubert}. Noting that near the zero of ${\cal D}$ (or the pole of the propagator), we can write
\begin{equation}
{\cal D}\xrightarrow{p^{2}\rightarrow M^{2}_{p}} Z (p^{2} - M^{2}_{p}),\label{gaugeind_3}
\end{equation}
where the coefficient $Z$ is related to the wave function normalization $Z_{2}^{-1}$, equation \eqref{gaugeind_2} leads to
\begin{equation}
\frac{\partial M_{p}}{\partial \phi_{(a)}} = 0.\label{gaugeind_4}
\end{equation}
Namely, the pole of the propagator or the physical mass of the fermion is independent of the three gauge fixing parameters $(\alpha, \beta, n^{\mu})$ in theories without infrared divergence and mass shell singularities at the pole. Furthermore, since the covariant, the axial and the Coulomb gauges correspond to specific values of these parameters which $M_{p}$ is independent of, the physical mass (or the location of the pole of the propagator) is the same in all three classes of gauges. This demonstrates the complete gauge independence of the fermion pole mass and it is worth emphasizing that this direct and simple demonstration of gauge invariance of the pole of the fermion propagator involves only three basic elements: choice of a general class of interpolating gauges, the Nielsen identity for the gauge variation of the fermion two point function for this gauge choice and the relation of the two point function to the denominator of the propagator.

We have explicitly verified the gauge independence of the pole of the fermion propagator, up to two loops, in the generalized axial gauge as follows. It is already known that at one loop, the pole is gauge parameter  independent (namely, $\widetilde{M} = M_{p}$ is gauge parameter independent up to one loop). Therefore, we need to concentrate only on the contributions at two loops. At two loops, the contributions to the self-energy are given by the diagrams Fig. \ref{fig1}. Using this in the expression for the denominator \eqref{axial_23} in the form,
$$
{\cal D} = - \frac{1}{2^{[n/2]}}\,{\rm Tr} \left(S^{-1}(p) S^{-1}(-p)\right),
$$
we can separate out the two loop contributions explicitly. Due to the transversality of the photon self-energy in diagram $(c)$ of Fig. \ref{fig1}, the gauge parameter ($(\beta, n^{\mu})$) dependent terms turn out to be proportional to $(p^{2}-m^{2})$ with a well behaved coefficient so that they vanish as $p^{2}\rightarrow m^{2}$ (note that since the amplitude is already of order two loops, $M_{p}^{2}$ can be set equal to $m^{2}$ in this order). Similarly the $\beta$ dependent terms in diagram $(d)$ as well as in the sum of the diagrams $(a)+(b)$ also have the same form with well behaved coefficients. Therefore, these also vanish at the pole and there is no $\beta$ dependence at all in the amplitude. The remaining $n^{\mu}$ dependent terms in the sum of the contributions from diagrams $(a)+(b)+(d)$ combine under the trace to be proportional to $(p^{2}-m^{2})$ (it is worth emphasizing that individual diagrams do not have this form). As a consequence, the complete two loop self-energy leads to  
\begin{equation}
{\cal D} = {\cal D}^{(Feynman)} + Q(p,m,n) (p^{2}-m^{2}),\label{gaugeind_5}
\end{equation}
where ${\cal D}^{(Feynman)}$ is manifestly gauge independent (calculated using $D_{\mu\nu}^{(Feynman)}$ in \eqref{axial_0c}) and $Q(p,m,n)$ is a complicated integral which is manifestly $n^{\mu}$ dependent (but has no $\beta$ dependence). It is clear from \eqref{gaugeind_5} that if there are no mass shell singularities in $Q(p,m,n)$ at the pole, then the gauge dependent terms in \eqref{gaugeind_5} vanish at the zero of the denominator, namely, as $p^{2}\rightarrow m^{2}$ (to this order) completely consistent with the conclusion following from the Nielsen identity in \eqref{gaugeind_2}. One can estimate the behavior of the coefficient function and near the mass shell it behaves like  $Q(p,m,n) \sim (p^{2}-m^{2})^{(n-4)/2}$ so that for $n\geq 4$ it is well behaved. However, for $n\leq 3$ mass shell singularities develop at the pole and one cannot conclude (as in \eqref{gaugeind_2}) that the pole of the propagator is gauge independent. We have seen this explicitly in an earlier study involving the massive Schwinger model \cite{schubert}. 

In fact, even in the Schwinger model (massless), where it is known that there is no infrared divergence, mass shell singularities do develop. For example, in the covariant gauge the fermion self-energy at one loop has the form (here we use the standard gauge fixing parameter $\xi = \alpha^{-2}$)
\begin{equation}
\Sigma^{(1)}_{(c)} (p) = \frac{\xi e^{2}}{2\pi p^{2}}\, p\sl,\label{gaugeind_6}
\end{equation}
and the singularity as $p^{2}\rightarrow 0$ is manifest in \eqref{gaugeind_6}. In this case, the pole of the fermion propagator is easily determined to be
\begin{equation}
M^{2}_{(c)} = \frac{\xi e^{2}}{2\pi},\label{gaugeind_7}
\end{equation}
and is manifestly gauge parameter dependent. Furthermore, one can calculate the one loop fermion self-energy in the generalized axial gauge which has the form
\begin{equation}
\Sigma^{(1)}_{(axial)} (p) = - \frac{e^{2}}{2\pi p_{\sc L}^{2}}\,p\sl_{\sc L},\label{gaugeind_8}
\end{equation}
leading to a fermion pole mass (in the perturbative regime $e^{2}\ll |p_{\sc L}^{2}|$)
\begin{equation}
M^{2}_{(axial)} = - \frac{e^{2}}{\pi}.\label{gaugeind_9}
\end{equation}
There are two things to note from \eqref{gaugeind_7} and \eqref{gaugeind_9}. First, the mass in the axial gauge is independent of the gauge fixing parameters $(\beta, n^{\mu})$ while that in the covariant gauge is manifestly gauge parameter dependent so that the two do not coincide in general. This is a simple example of how a gauge parameter independent pole within one class of gauges (in this case the axial gauge) does not automatically imply that it will be the same in all classes of gauges. Second, because the poles in the two gauges do not coincide, even if the pole in the axial gauge is gauge parameter independent, it will be incorrect to conclude that it represents a physical mass. In fact, we see from \eqref{gaugeind_9} that this mass is purely imaginary and, therefore, completely unphysical. 

\section{Conclusion}

In this paper we have studied the question of complete gauge independence of the fermion pole mass. This involves showing the gauge parameter independence of the pole within a given class of gauges (such as covariant or axial or Coulomb) as well as showing that the pole has the same value in all of these three classes of gauges. The demonstration of complete gauge independence is achieved in a simple manner by using three basic ingredients. First, we choose a general class of gauges which interpolate between the covariant, the axial and the Coulomb gauges for different values of the gauge parameters. We derive the Nielsen identity describing the gauge parameter variation of the fermion two point function in this class of  interpolating gauges. And we relate the denominator of the fermion propagator to the two point function so that the gauge parameter variation of the denominator (or the pole of the propagator) can be studied directly  using the Nielsen identity for the fermion two point function. With these three basic ingredients we are able to show in a simple manner that, when there are no infrared divergences and mass shell singularities at the pole, the pole of the fermion propagator is gauge independent in the complete sense. The presence of mass shell singularities can invalidate such a proof and this is pointed out with a simple example of the Schwinger model where the pole mass is manifestly gauge parameter dependent in the covariant gauge while it is independent of gauge parameters in the generalized axial gauges. The pole masses in different classes of gauges do not coincide in this case and, therefore, it would be wrong to conclude from a study of the pole in the axial gauge (where it is gauge parameter independent) that the pole represents a physical mass. In fact, the pole mass in this case (in the axial gauge) turns out to be purely imaginary and, therefore, unphysical. Physically, of course, this can be understood from the fact that in $1+1$ dimensions, the Coulomb potential increases linearly with distance leading to confinement and preventing any physical fermions in the asymptotic states.
   
\bigskip

\noindent{\bf Acknowledgments}
\medskip

 A. D. would like to thank the Departamento de F\'{i}sica Matematica for hospitality where this work was done. This work was supported in part by USP, by CNPq and by FAPESP.

\appendix
\section{Quantum Chromodynamics}
 
In this appendix, we will briefly indicate how the derivation of the Nielsen identity generalizes to Quantum Chromodynamics in a straightforward manner. For a non-Abelian theory based on the gauge group $SU(N)$ the invariant Lagrangian density is given by
\begin{equation}
{\cal L}_{\rm inv} = - \frac{1}{4}\, F_{\mu\nu}^{a} F^{\mu\nu, a} + \overline{\psi}_{i} (i D\!\sl -m)\psi_{i},\label{app_1}
\end{equation}
where $a=1,2,\cdots , N^{2}-1$ and $i=1,2,\cdots , N$ denote the color indices. The covariant derivative for the fermion is defined to be
\begin{equation}
D_{\mu}\psi_{i} = \partial_{\mu}\psi_{i} + i gA_{\mu}^{a} (T^{a})_{ij} \psi_{j},\label{app_2}
\end{equation}
where $T^{a}$ denotes the (Hermitian) generators of the group in the fundamental representation and $g$ denotes the coupling constant. One can generalize the interpolating gauge fixing in this case to be given by
\begin{equation}
{\cal L}_{\rm\sc GF} = -\frac{1}{2} \left(\Lambda\cdot A^{a}\right)^{2} = \frac{1}{2}\,F^{a}F^{a} + (\Lambda^{\mu}F^{a}) A_{\mu}^{a}.\label{app_3}
\end{equation}
where $\Lambda^{\mu} = \Lambda^{\mu}(\partial)$ is given in \eqref{nielsen_3}. 

The ghost Lagrangian density similarly generalizes to
\begin{equation}
{\cal L}_{\rm ghost} = (\Lambda^{\mu}\overline{c}^{a}) D_{\mu}c^{a},\label{app_4}
\end{equation}
where the covariant derivative in the adjoint representation is defined as
\begin{equation}
D_{\mu}c^{a} = \partial_{\mu}c^{a} - g f^{abc} A_{\mu}^{b} c^{c},\label{app_5}
\end{equation}
with $f^{abc}$ representing the structure constants of the group. The Lagrangian density ${\cal L}_{\rm inv} + {\cal L}_{\rm\sc GF} + {\cal L}_{\rm ghost}$ is manifestly invariant under the standard (nilpotent) BRST transformations (see also \eqref{nielsen_7})
\begin{align}
\delta A_{\mu}^{a} & = \omega (D_{\mu}c)^{a},\notag\\
\delta \psi_{i} & = -ig\omega c^{a} (T^{a})_{ij}\psi_{j},\notag\\
\delta \overline{\psi}_{i} & = - ig\omega \overline{\psi}_{j} (T^{a})_{ji} c^{a},\notag\\
\delta c^{a} & = \frac{g\omega}{2}\, f^{abc} c^{b}c^{c},\notag\\
\delta\overline{c}^{a} & = -\omega F^{a},\notag\\
\delta F^{a} & = 0,\label{app_6}
\end{align}
where $\omega$ denotes a constant Grassmann parameter. The important thing to note here is that in the non-Abelian theory, the quadratic part of the Lagrangian density is diagonal in the color space so that the two point functions have the same structure as in the Abelian theory except for diagonal color factors such as $\delta^{ab}$ or $\delta_{ij}$. Therefore, much of our discussion goes through with minimal change in this case. For example, the denominator for the fermion propagator can now be related to the fermion two point function as in \eqref{axial_23} with the \lq\lq Tr" representing a trace over the spinor as well as the color indices and the normalization factor will be correspondingly different. (These are minor generalizations which have no effect on the final result.)

Since the BRST transformations \eqref{app_6} are somewhat different (there are more composite variations), so are the source terms necessary to derive the Nielsen identity. In this case, the source Lagrangian density is given by
\begin{widetext}
\begin{align}
{\cal L}_{\rm source} & = J^{\mu, a}A_{\mu}^{a} + J^{a} F^{a} + i (\overline{\chi}_{i}\psi_{i} - \overline{\psi}_{i}\chi_{i} )+ i(\overline{\eta}^{a} c^{a} - \overline{c}^{a}\eta^{a}) 
 + K^{\mu a} D_{\mu}c^{a} + ig (\overline{M}_{i} c^{a}(T^{a})_{ij}\psi_{j} - \overline{\psi}_{i}(T^{a})_{ij}c^{a}M_{j})\notag\\
 & \ \ + K^{a} \left(\frac{g}{2} f^{abc} c^{b}c^{c}\right) + H_{(\alpha)} (\partial^{\mu}\overline{c}^{a}) A_{\mu}^{a} + H_{(\beta)} (\partial_{\sc L}^{\mu}\overline{c}^{a}) A_{\mu}^{a} + \beta H_{(n) \mu} (N^{\mu\nu}\overline{c}^{a}) A_{\nu}^{a},\label{app_7}
\end{align}
\end{widetext}
where $N^{\mu\nu}$ is defined in \eqref{nielsen_9}. With this we can define the total Lagrangian density for the theory to be
\begin{equation}
{\cal L}_{\rm\sc TOT} = {\cal L}_{\rm inv} + {\cal L}_{\rm\sc GF} + {\cal L}_{\rm ghost} + {\cal L}_{\rm source}.\label{app_8}
\end{equation}
One can now follow through the steps in \eqref{nielsen_12}-\eqref{nielsen_16} to obtain the Master identity for the gauge parameter variation of the generating functional for the connected Green's functions (the Dirac spinor indices have been suppressed on the right hand side for simplicity)
\begin{widetext}
\begin{align}
\frac{\partial W}{\partial\phi_{(a)}} & = i\int d^{n}x d^{n}y \left(i J^{\mu a} (y) \frac{\delta^{2}W}{\delta H_{(a)} (x)\delta K^{\mu a} (y)} + \overline{\chi}_{i}(y) \frac{\delta^{2}W}{\delta\overline{M}_{i}(y) \delta H_{(a)}(x)} + \frac{\delta^{2}W}{\delta H_{(a)}(x)\delta M_{i}(y)} \chi_{i}(y)\right.\notag\\
& \qquad\qquad\qquad  \left. - \frac{\delta^{2}W}{\delta H_{(a)}(x)\delta J^{a}(y)} \eta^{a}(y) - \overline{\eta}^{a}(y) \frac{\delta^{2}W}{\delta H_{(a)}(x)\delta K^{a}(y)}\right).\label{app_9}
\end{align}
\end{widetext}
This can be compared with the identity for the Abelian case in \eqref{nielsen_16}. 

The Legendre transformation with respect to the standard sources for the field variables (see \eqref{nielsen_17}) takes us to the Master identity for the gauge parameter variation of the effective action which has the form (we are suppressing the Dirac spinor indices on the right hand side for simplicity)
\begin{widetext}
\begin{align}
\frac{\partial\Gamma}{\partial\phi_{(a)}} & = \int d^{n}x d^{n} y\left[\frac{\delta\Gamma}{\delta A_{\mu}^{a} (y)} \frac{\delta^{2}\Gamma}{\delta H_{(a)}(x) \delta K^{\mu a} (y)} + \frac{\delta\Gamma}{\delta \psi_{i}(y)}\frac{\delta^{2}\Gamma}{\delta \overline{M}_{i} (y) \delta H_{(a)} (x)} + \frac{\delta^{2}\Gamma}{\delta H_{(a)}(x) \delta M_{i}(y)} \frac{\delta\Gamma}{\delta \overline{\psi}_{i}(y)}\right.\notag\\
&\qquad\qquad\qquad \left. - \frac{\delta\Gamma}{\delta c^{a} (y)} \frac{\delta^{2}\Gamma}{\delta H_{(a)} (x) \delta K^{a}(y)}\right],\label{app_10}
\end{align}
\end{widetext}
where all the fields (and the sources $M_{i}, \overline{M}_{i}$) have been set to zero. (This can be compared with \eqref{nielsen_18} for the Abelian theory.) Taking the functional derivative with respect $\frac{\delta^{2}}{\delta\psi_{j,\beta} (y) \delta\overline{\psi}_{i,\alpha} (x)}$, we can now obtain the gauge parameter variation of the fermion two point function which, in the matrix form (for simplicity), is given by
\begin{align}
\frac{\partial S^{-1} (x-y)}{\partial\phi_{(a)}} & = \int d^{n}z d^{n} w\left({\cal F}^{(a)} (x,z,w) S^{-1} (w-y)\right.\notag\\
&\qquad\quad \left. + S^{-1}(x-w) {\cal G}^{(a)} (w,z,y)\right),\label{app_11}
\end{align}
where we have identified (here we put back all the indices)
\begin{align}
& S^{-1}_{ij,\alpha\beta} (x-y) = \frac{\delta^{2}\Gamma}{\delta\psi_{j,\beta}(y) \delta\overline{\psi}_{i,\alpha}(x)},\notag\\
& {\cal F}^{(a)}_{ik,\alpha\gamma}(x,z,w) = \frac{\delta^{3}\Gamma}{\delta\overline{\psi}_{i,\alpha}(x)\delta H_{(a)}(z)\delta M_{k,\gamma}(w)},\notag\\
& {\cal G}^{(a)}_{kj,\gamma\beta} (w,z,y) = \frac{\delta^{3}\Gamma}{\delta\overline{M}_{k,\gamma}(w) \delta H_{(a)}(z)\delta \psi_{j,\beta}(y)}.\label{app_12}
\end{align}
In momentum space this relation takes the simple (matrix) form
\begin{equation}
\frac{\partial S^{-1} (p)}{\partial\phi_{(a)}} =  {\cal F}^{(a)} (p) S^{-1}(p) + S^{-1} (p) {\cal G}^{(a)}(p),\label{app_13}
\end{equation}
where, as in \eqref{nielsen_22}, we have identified the matrices 
\begin{align}
{\cal F}^{(a)} (p) & = {\cal F}^{(a)} (-p,0,p),\notag\\
{\cal G}^{(a)} (p) & = {\cal G}^{(a)} (-p,0,p).\label{app_14}
\end{align}
We note that in the matrix form the gauge parameter variation of the fermion two point function \eqref{app_13}  has the same form as in \eqref{nielsen_21}, the difference being that in the non-Abelian case, the matrices are matrices in the color as well as spinor space.

As we have mentioned earlier, the fermion two point function as well as the propagator are diagonal in the color space. So, \eqref{axial_22} still holds with the understanding that the identity matrix is a matrix in the color as well as the spinor space. This determines (see \eqref{axial_23}) the denominator of the propagator to be given by
\begin{align}
{\cal D} & = - \frac{1}{2^{[n/2]} N}\,{\rm Tr} \left(S^{-1}(p) ({\cal C} (S^{-1}(p))^{T} {\cal C}^{-1})\right)\notag\\
& = - \frac{1}{2^{[n/2]} N}\, {\rm Tr} \left(S^{-1}(p) S^{-1}(-p)\right),\label{app_15}
\end{align}
where the trace is over color as well as spinor indices. We can now determine the variation of the denominator with respect to the gauge parameters using \eqref{app_13} (exactly as was done in \eqref{gaugeind_1}) to be given by
\begin{equation}
\frac{\partial{\cal D}}{\partial\phi_{(a)}} = 2 {\cal D}\,{\rm Tr} \left({\cal F}^{(a)} (p) + {\cal G}^{(a)} (p)\right),\label{app_16}
\end{equation}
and the gauge independence of the pole of the propagator follows exactly as discussed in section {\bf 4}.

\end{document}